\documentclass[11pt]{article}

\usepackage{hyperref}
\usepackage{amsmath}
\usepackage{amsthm}
\usepackage{amssymb}
\usepackage{algorithm}
\usepackage{algpseudocode}

\newtheorem{theorem}{Theorem}

\newtheorem{proposition}{Proposition}

\title{Efficient, traceable, and numerically error-free implementation
  of the MMS voting rule}

\author{Luis S\'anchez-Fern\'andez
\\
Universidad Carlos III de Madrid, Spain \\
luiss@it.uc3m.es
}

\date{July 14, 2026}

\begin{document}

\maketitle

\begin{abstract}
  
  We propose an alternative algorithm to compute the MMS voting
  rule. Instead of using linear programming, in this new algorithm, the
  maximin support value of a committee is computed using a sequence of
  maximum flow problems.

\end{abstract}

\section{Introduction}

The maximin support method (MMS) is an extension of the well-known
D'Hondt method (or Jefferson method) for approval-based multi-winner
elections proposed by~\cite{sanchez2024maximin}. It possesses several
nice axiomatic properties related to representation and
monotonicity. It satisfies proportional justified
representation~\cite{pjr-aij}, weak support
monotonicity~\cite{sanchez2019monotonicity}, and committee
monotonicity. Moreover,~\cite{CeSt21a} proved a worst-case
$2$-approximation of MMS to the rule
max-Phragm\'en~\cite{2016arXiv161108826J,brill:phragmen} proposed in
the nineteenth century by the Swedish mathematician Lars Edvard
Phragm\'en.

\cite{sanchez2024maximin} also proved that the MMS rule can be computed in polynomial time. In their proof,~\cite{sanchez2024maximin}
used a linear program to compute the maximin support values (see the
definition of MMS in Section~\ref{sec:MMS}).

Using linear programming to compute the outcome of the MMS rule
has several disadvantages. First, although linear programs can be
solved in polynomial time, their worst-case complexity is high. The
best-known result for solving linear programs is due
to~\cite {karmarkar84}, which proved a complexity upper bound of
$\mathcal{O}(n^{3.5} L)$, where $n$ is the number of variables in the
linear program, and $L$ is the number of bits required to represent the
absolute values of all the non-zero coefficients contained in the
linear program. Second, although commercial linear programming solvers
are well-established, highly reliable systems commonly used
in a broad set of practical optimization problems, their use in the
delicate process of computing the outcome of an election can be the
object of a certain level of mistrust. Third, commercial linear
programming solvers use floating-point arithmetic to compute the
solution of linear programs, and therefore, they do not provide an
exact solution. This may induce problems, for instance, in determining
when a tie happens.

In this note, we propose an alternative way to compute the MMS voting
rule. Instead of using a linear program, we will use a sequence of
maximum flow problems to compute the maximin support values (at most
$k$ for each maximin support value, where $k$ is the size of the
committee).

This alternative algorithm solves all the problems mentioned above
related to using linear programming. First, since maximum flow
problems can be computed efficiently~\cite{goldberg2014efficient}, their
use can improve the computation time of the MMS rule. Second,
algorithms for computing maximum flow problems are publicly available,
so anybody can implement them and avoid the need to use 'black-box'
software tools. Third, the algorithm proposed in this note only uses
integer arithmetic; therefore, this algorithm obtains an exact
solution and is free of numerical errors.

\section{Preliminaries}

\subsection{Approval-based multi-winner elections}

We model an approval-based multi-winner election as a tuple
$(N,C,\mathcal{A})$, where $N= \{1, \ldots, n\}$ is a set of voters
and $C$ is a nonempty set of candidates. $\mathcal{A}= (A_1, \ldots,
A_n)$ is the ballot profile. For each voter $i$ in $N$, $A_i \subseteq
C$ is the set of candidates that voter $i$ approves. 

An approval-based multi-winner voting rule $R$ receives as input an
election $(N,C,\mathcal{A})$, and a target committee size $k$ (we
assume $1 \leq k \leq |C|$), and outputs a committee (a subset of the
candidates) $W$ of size $k$. We also assume that voting rules are
resolute.

In many cases, we will be interested only in a particular subset $S$ of
the candidates. Given an election $\mathcal{E}= (N,C,\mathcal{A})$ and
a subset $S \subseteq C$, we say that $\mathcal{E}|_S=
(N_S,S,\mathcal{A}_S)$ is the {\it restriction} of election
$(N,C,\mathcal{A})$ to the candidates subset $S$, where $N_S= \{i \in
N: A_i \cap S \ne \emptyset\}$, $\mathcal{A}_S= (A_{iS}: i \in N_S)$,
and $A_{iS}= A_i \cap S$, for each $i$ in $N_S$. In words, the
restriction of an election to a subset $S$ of the candidates considers only
voters that approve some candidates in $S$ and, for such voters,
considers only the candidates approved by them that belong to $S$.

\subsection{Maximum flow problems}

A flow network is composed of a directed graph $G= (V,E)$ a source
vertex $s$ in $V$, a sink vertex $t$ in $V$, and integral capacity
function $u: E \rightarrow \mathbb{N}^+$, where for each edge $e$ in
$E$, $u(e)$ is a positive integer that represents the maximum flow
capacity of edge $e$.

A flow is a function $f: E \rightarrow \mathbb{N}$ satisfying (i)
$f(e) \leq u(e)$ for each edge $e$ in $E$; and (ii) $\sum_{(j,k) \in E}
f(j,k) - \sum_{(i,j) \in E} f(i,j)= 0$, for each node $j$ in $V
\setminus \{s,t\}$.

The value $|f|$ of a flow $f$ is the net flow into the sink: $|f|=
\sum_{(v,t) \in E} f(v,t)$. The maximum flow problem consists of
finding a flow with a maximum flow value.

\section{Definition of MMS}
\label{sec:MMS}

This section describes the MMS voting rule. All the concepts and
definitions included in this section are taken
from~\cite{sanchez2024maximin}.

\subsection{Maximin support value}

Given an approval-based multi-winner election $\mathcal{E}=
(N,C,\mathcal{A})$ and a candidates subset $S$, the family
$\mathcal{F}_{\mathcal{E},S}$ of vote assignments for
$(\mathcal{E},S)$ is the set of all functions $\alpha: (N \times S)
\rightarrow [0,1]$ that distribute voter support only among the
candidates in $S$.

Each vote assignments $\alpha$ in $\mathcal{F}_{\mathcal{E},S}$ must
fulfill the following conditions: (i) $\alpha(i,c) = 0$ for each candidate
$c$ in $S \setminus A_i$; and (ii) $\sum_{c \in A_i \cap S} \alpha(i,c)= 1$
for each voter $i$ such that $A_i \cap S \ne \emptyset$.

A vote assignment $\alpha$ associates a value of support with each candidate
in $S$. Formally, $\mathit{supp}_\alpha(c)= \sum_{i \in N} \alpha(i,c)$, for
each vote assignment $\alpha$ in $\mathcal{F}_{\mathcal{E},S}$, and each
candidate $c$ in $S$.

Given an approval-based multi-winner election $\mathcal{E}=
(N,C,\mathcal{A})$ and a candidates subset $S$, the maximin support
value of $(\mathcal{E},S)$ is the supremum over all possible vote
assignments $\alpha$ in $\mathcal{F}_{\mathcal{E},S}$ of the support of the
least supported candidate under vote assignment $\alpha$:

\begin{equation}
  \mathit{maximin}(\mathcal{E},S)= \max_{\alpha \in \mathcal{F}_{\mathcal{E},S}}
    \min_{c \in S} \mathit{supp}_\alpha(c)
\end{equation}

For a given approval-based multi-winner election $\mathcal{E}=
(N,C,\mathcal{A})$ and a candidates subset $S$, let
$\mathcal{F}_{\mathcal{E},S}^{\textrm{opt}}$ be the set of optimal
vote assignments that maximize the support of the least supported
candidate. $\mathit{supp}_\alpha(c) \geq \mathit{maximin}(\mathcal{E},S)$
for each vote assignment $\alpha$ in
$\mathcal{F}_{\mathcal{E},S}^{\textrm{opt}}$, and each $c$ in $S$.

\subsection{The MMS voting rule}

The MMS voting rule is an iterative algorithm. It starts with an empty
set of winners $W$. At each iteration, the candidate $c$ in $C
\setminus W$ added to the set of winners is the one with the
highest support under the condition that the support of the least
supported candidate is maximized. The process is repeated until the
size of $W$ equals the target committee size $k$.

\cite{sanchez2024maximin} proved that the support of candidate $c$
under any vote assignment that maximizes the support of the least
supported candidate coincides with $\mathit{maximin}(\mathcal{E},W
\cup \{c\})$. This implies that selecting the candidate with the highest
support under the condition that the support of the least supported
candidate is maximized is equivalent to selecting the candidate that
maximizes the maximin support value.

\section{Computing the maximin support values with maximum flow problems}

Given an approval-based multi-winner election $\mathcal{E}=
(N,C,\mathcal{A})$ and a subset $S$ of the candidates, we define an associated
flow network as follows. First, we compute the restriction of
$(N,C,\mathcal{A})$ to $S$= $(N_S,S,\mathcal{A}_S)$. Then, the set of
nodes $V$ in the flow network is $N_S \cup S \cup \{s, t\}$. The set
of edges $E$ consists of (i) an edge $(s,i)$ for each voter $i$ in
$N_S$ with capacity $u(s,i)= 1$; (ii) an edge $(i,c)$ for each voter
$i$ in $N_S$ and each candidate $c$ in $A_{iS}$ with capacity $u(i,c)=
1$; and (iii) an edge $(c,t)$ for each candidate $c$ in $S$ with
capacity $u(c,t)= \frac{|N_S|}{|S|}$ (this value may not be integral; we
will fix this later). Any flow in this flow network can freely distribute
the unit of flow available from the source $s$ to each voter
$i$ in $N_S$ among the candidates in $S$ that such voter
approves. However, each candidate can send at most $\frac{|N_S|}{|S|}$
units of flow to the sink $t$.

Now, if the maximum flow value of this flow network is equal to
$|N_S|$, then we are done: there exists a flow such that each
candidate receives $\frac{|N_S|}{|S|}$ units of flow which in turns
means that it is possible to distribute the voter support so that each
candidate in $S$ has a support of $\frac{|N_S|}{|S|}$ and therefore
$\mathit{maximin}(\mathcal{E}, S)= \frac{|N_S|}{|S|}$.

For the case where the maximum flow value of this flow network is
strictly smaller than $|N_S|$, first we will prove the following result.


\begin{theorem}
  \label{th:kernel}
For any approval-based multi-winner election $\mathcal{E}=
(N,C,\mathcal{A})$ and any non-empty candidate subset $S$, a non-empty
subset $K$ of $S$ exists satisfying:

\begin{enumerate}

\item

$\mathit{supp}_\alpha(c)= \mathit{maximin}(\mathcal{E},S)$
for each vote assignment $\alpha$ in
$\mathcal{F}_{\mathcal{E},S}^{\textrm{opt}}$, and each $c$ in $K$; and

\item
 
  $\mathit{maximin}(\mathcal{E},S)=
  \frac{|\{i \in N: A_i \cap K \ne \emptyset\}|}{|K|}$ 
  
\end{enumerate}  
.

\end{theorem}

We will use the following
result\footnote{Proposition~\ref{prop:markus} is stated as Lemma~A.1
in~\cite{sanchez2024maximin}.} by~\cite{sanchez2024maximin} in the
proof of Theorem~\ref{th:kernel}.

\begin{proposition}[\cite{sanchez2024maximin}]
  \label{prop:markus}

Let $\mathcal{E}= (N,C,\mathcal{A})$ be an approval-based multi-winner election, $S \subseteq C$ a nonempty set of candidates and $\alpha \in \mathcal{F}_{\mathcal{E},S}^{\textrm{opt}}$ an optimal vote assignment for $(\mathcal{E}, S)$. If exists a candidate $\hat{c} \in S$ such that $\mathit{supp}_\alpha(\hat{c}) > \mathit{maximin}(\mathcal{E},S)$, then $\mathit{maximin}(\mathcal{E},S)= \mathit{maximin}(\mathcal{E},S\setminus\{\hat{c}\})$.
  
\end{proposition}  

\begin{proof}[Proof of Theorem \ref{th:kernel}]

  Pick an arbitrary optimal vote assignment $\alpha \in \mathcal{F}_{\mathcal{E},S}^{\textrm{opt}}$. If for some candidate $\hat{c} \in S$ it holds that $\mathit{supp}_\alpha(\hat{c}) > \mathit{maximin}(\mathcal{E},S)$, remove candidate $\hat{c}$ from $S$. Repeat this process iteratively until, at some iteration, it holds that $\mathit{supp}_{\alpha_K}(c) = \mathit{maximin}(\mathcal{E},K)$ for each candidate $c \in K$, where $K$ is the subset of $S$ obtained with this iterative removal of candidates and $\alpha_K \in \mathcal{F}_{\mathcal{E},K}^{\textrm{opt}}$. If $\mathit{supp}_\alpha(c) = \mathit{maximin}(\mathcal{E},S)$ for each candidate $c$ in $S$ simply set $K= S$.

  By also applying iteratively Proposition~\ref{prop:markus}, it holds that $\mathit{maximin}(\mathcal{E},K)= \mathit{maximin}(\mathcal{E},S)$, because at each iteration we remove a candidate with support greater than the maximin support value, and thus, the maximin support value does not change. Observe also that $K$ is nonempty, because by the definitions of maximin support value and optimal vote assignment, at each iteration, some candidate must exist such that its support is the maximin support value. Thus, such a candidate cannot be removed.

  Now, by the definitions of vote assignment and support, we have:

  \begin{displaymath}
    \sum_{c \in K} \mathit{supp}_{\alpha_K}(c)= \sum_{c \in K} \sum_{i \in N} \alpha_K(i,c)= \sum_{i \in N} \sum_{c \in K} \alpha_K(i,c) = |\{i \in N: A_i \cap K \ne \emptyset\}|
  \end{displaymath}

  The last equality comes from the fact that $\sum_{c \in K} \alpha_K(i,c)$ is equal to $1$ if voter $i$ approves some candidate in $K$ and $0$ otherwise.

  But since $\mathit{supp}_{\alpha_K}(c) = \mathit{maximin}(\mathcal{E},K) = \mathit{maximin}(\mathcal{E},S)$ for each candidate $c \in K$, we also have

  \begin{displaymath}
    \sum_{c \in K} \mathit{supp}_{\alpha_K}(c)= |K| \mathit{maximin}(\mathcal{E},S),
  \end{displaymath}

  and therefore,

  \begin{displaymath}
\mathit{maximin}(\mathcal{E},S)=
\frac{|\{i \in N: A_i \cap K \ne \emptyset\}|}{|K|}
  \end{displaymath}

  This proves the second part of the theorem.

  For the first part of the theorem observe that by definition of optimal vote assignment it must be $\mathit{supp}_{\alpha}(c) \geq \mathit{maximin}(\mathcal{E},S)$ for each candidate $c$ in $K$ and each $\alpha \in \mathcal{F}_{\mathcal{E},S}^{\textrm{opt}}$, and since the number of voters that approve some candidate in $K$ is equal to $|K| \mathit{maximin}(\mathcal{E},S)$, then it has to be $\mathit{supp}_{\alpha}(c) = \mathit{maximin}(\mathcal{E},S)$ for each candidate $c$ in $K$ and each $\alpha \in \mathcal{F}_{\mathcal{E},S}^{\textrm{opt}}$.

\end{proof}

If the maximum flow value of the flow network is strictly smaller than
$|N_S|$, then for any flow $f$ with maximum flow value a voter $i$
must exist such that $\sum_{(i,c) \in E} f(i,c) < 1$ (if such voter
does not exist, then the maximum flow value should be equal to
$|N_S|$). Any candidate approved by voter $i$ cannot belong to $K$,
because by Theorem~\ref{th:kernel}, we know that it is possible
to assign the votes of all voters who approve some candidate in $K$
only between the candidates in $K$, so that all candidates in $K$
receive a support equal to $\mathit{maximin}(\mathcal{E},S)$ which is
smaller than the capacity of the edges that go from the candidates to
the sink (that it is equal to $\frac{|N_S|}{|S|}$). Therefore, for any
flow $f$ with maximum flow value and any voter $i'$ that approves some
candidate in $K$, it must be $\sum_{(i',c) \in E} f(i',c) =
1$.\footnote{If some candidates in $S$ are not approved by any voter
in $\mathcal{E}$, then every voter will be removed from $N_S$ at some
iteration. This is because while some voters remain in $N_S$, it is
not possible that each candidate sends $\frac{|N_S|}{|S|}$ units of
flow to the sink because the candidates not approved by any voter cannot send any flow to the sink. We will end with $N_S=
\emptyset$ (all voters are successively removed from $N_S$), and $S$
would contain only those candidates that are not approved by any
voter. As expected, it would be $\mathit{maximin}(\mathcal{E},S)=
\frac{|N_S|}{|S|}= \frac{|\emptyset|}{|S|}= 0$.}

We remove from $S$ all the candidates that are approved by each
voter $i$ such that $\sum_{(i,c) \in E} f(i,c) < 1$, and repeat the
process until we get a flow network with value $|N_S|$.

Finally, to avoid the problem that the capacities of the edges from
the candidates to the sink may not be integral, we multiply the
capacities of all edges by $|S|$, and adapt the algorithm accordingly.

The new method to compute the maximin support values is defined in
Algorithm~\ref{alg:mmsflow}.

\begin{algorithm}[htb]
\caption{Computation of the maximin support value\label{alg:mmsflow}}
{\bf Input}: an approval-based multi-winner election $\mathcal{E}= (N,C,\mathcal{A})$\\
and a non-empty subset $S$ of $C$.\\         
{\bf Output}: the maximin support value of $(\mathcal{E},S)$.
\algblockdefx[ForEach]{ForEach}{EndForEach}[1]
{\textbf{foreach} #1 \textbf{do}}{\textbf{end foreach}} 
\begin{algorithmic}
\State $(N_S,S,\mathcal{A}_S) \gets (N, C, \mathcal{A})|_S$
\State $e \gets$ {\bf false}
\Repeat
  \State $V \gets N_S \cup S \cup \{s, t\}$
  \State $E \gets \emptyset$
  \ForEach{$i \in N_S$}
    \State $E \gets E \cup \{(s,i)\}$
    \State $u(s,i) \gets |S|$
    \ForEach{$c \in A_{iS}$}
      \State $E \gets E \cup \{(i,c)\}$
      \State $u(i,c) \gets |S|$
    \EndForEach
  \EndForEach
  \ForEach{$c \in S$}
    \State $E \gets E \cup \{(c,t)\}$
    \State $u(c,t) \gets |N_S|$
  \EndForEach
  \State Compute a flow $f$ with maximum flow value for $(G=(V,E),u)$
  \State $v \gets \sum_{(c,t) \in E} f(c,t)$
  \If{$v < |N_S||S|$}
    \State $\mathit{Drop} \gets \emptyset$
    \ForEach{$i \in N_S$}
      \If{$\sum_{(i,c) \in E} f(i,c) < |S|$}
        \State $\mathit{Drop} \gets \mathit{Drop} \cup A_{iS}$
      \EndIf
    \EndForEach
    \State $S' \gets S \setminus \mathit{Drop}$
    \State $(N_S,S,\mathcal{A}_S) \gets (N_S, S, \mathcal{A}_S)|_{S'}$
  \Else
    \State $e \gets$ {\bf true}
  \EndIf  
\Until{$e$}
\State \Return{$\frac{|N_S|}{|S|}$}  
\end{algorithmic}
\end{algorithm}

\bibliographystyle{ACM-Reference-Format}
\bibliography{dhondt}

@article{goldberg2014efficient,
  title={Efficient maximum flow algorithms},
  author={Goldberg, Andrew V and Tarjan, Robert E},
  journal={Communications of the ACM},
  volume={57},
  number={8},
  pages={82--89},
  year={2014},
  publisher={ACM New York, NY, USA}
}

@string{aamas19 = "Proceedings of the 18th International Conference on Autonomous Agents and Multiagent Systems (AAMAS-2019)"}

@article{sanchez2024maximin,
  title={The {Maximin Support Method}: {A}n extension of the {D'Hondt} method to approval-based multiwinner elections},
  author={S{\'a}nchez-Fern{\'a}ndez, Luis and Fern{\'a}ndez-Garc{\'i}a, Norberto and Fisteus, Jes{\'u}s A. and Brill, Markus},
  journal={Mathematical Programming},
year={2024},
volume={203},
pages= {107--134},
issn={1436-4646},
doi={10.1007/s10107-022-01805-8},
url={https://doi.org/10.1007/s10107-022-01805-8}
}

@inproceedings{sanchez2019monotonicity,
  title={Monotonicity Axioms in Approval-based Multi-winner Voting Rules},
  author={S{\'a}nchez-Fern{\'a}ndez, Luis and Fisteus, Jes{\'u}s A},
  booktitle=aamas19,
  pages={485--493},
  year={2019},
  publisher={IFAAMAS}
}

@inproceedings{CeSt21a,
author = {Cevallos, Alfonso and Stewart, Alistair},
title = {A Verifiably Secure and Proportional Committee Election Rule},
year = {2021},
booktitle = {Proceedings of the 3rd ACM Conference on Advances in Financial Technologies (AFT-2021)},
pages = {29--42},
publisher = {Association for Computing Machinery},
address = {New York, NY, USA}
}

@techreport{2016arXiv161108826J,
   author = {{Janson}, S.},
title = {Phragm{\'e}n's and {T}hiele's election methods},
archivePrefix = "arXiv",
 primaryClass = "math.HO",
     year = 2016,
    month = nov,
    institution = {arXiv:1611.08826 [math.HO]},
}

@article{brill:phragmen,
  title={Phragm{\'e}n’s voting methods and justified representation},
  author={Brill, Markus and Freeman, Rupert and Janson, Svante and Lackner, Martin},
  journal={Mathematical Programming},
  pages={47--76},
  year={2024},
  volume=203,
  publisher={Springer}
}

@ARTICLE{karmarkar84,
  AUTHOR = {N. Karmarkar},
  TITLE = {A new polynomial-time algorithm for linear programming},
  JOURNAL = {Combinatorica},
  YEAR = {1984},
  VOLUME = {4},
  PAGES = {373--395},
}

@article{pjr-aij,
	Author = {L. S{\'a}nchez-Fern{\'a}ndez and E. Elkind and M. Lackner and N. Fern{\'a}ndez and J. A. Fisteus and P. {Basanta Val} and P. Skowron},
title = {Proportional justified representation},
journal = {Artificial Intelligence},
volume = {353},
pages = {104503},
year = {2026},
issn = {0004-3702},
doi = {https://doi.org/10.1016/j.artint.2026.104503},
url = {https://www.sciencedirect.com/science/article/pii/S0004370226000299}
	}

\end{document}